\documentstyle[12pt,epsf]{article}
\newbox\mystrutbox
\setbox\mystrutbox=\hbox{\vrule height 17pt depth 7pt width 0pt}
\def\mystr{\relax\ifmmode\copy\mystrutbox\else\unhcopy\mystrutbox\fi}
\begin{document}
\def\mediumspace{\baselineskip 18pt \lineskip 6pt \parskip 3pt plus 5pt}
\def\doublespace{\baselineskip 24pt \lineskip 10pt \parskip 5pt plus 10pt}

\begin{flushright}
DOE-40200-014\\
CPP-14\\
MSUHEP 93/08
\end{flushright}

\doublespace
\vspace{12pt}
\begin{center}
 \large Photon neutrino scattering
\normalsize

\vspace{36pt}
Duane A. Dicus \\

\vspace{12pt}
{\it Center for Particle Physics and Department of Physics\\
University of Texas, Austin, Texas 78712}

\vspace{24pt}
Wayne W. Repko \\

\vspace{12pt}
{\it Department of Physics and Astronomy \\
Michigan State University, East Lansing, Michigan 48824}

\vspace{.8 truein}
{\large Abstract}
\end{center}

{\textwidth 6.0 truein The cross section for photon neutrino scattering is
calculated in the standard model assuming that the neutrino is massless and
that
the center of mass energy is small compared to any charged lepton mass.
Although
the scattered photons can acquire a (parity violating) circular polarization of
order unity, the cross section in this limit is highly suppressed.}

\pagebreak

\setcounter{page}{1}
\noindent{\large\sc 1.~Introduction}

Low energy neutrino photon scattering, e.g., $\gamma\gamma\rightarrow
\nu\bar{\nu}$ could be
of interest in astrophysics.  Unfortunately this interaction is known to be
highly suppressed. In the four-Fermi limit of the standard model it is exactly
zero\,\cite{gellmann} since a vector (or axial vector) cannot couple to two
massless vectors\,\cite{yang}. Thus the interaction is smaller than a naive
counting of couplings would suggest by factors of $\omega/M_W$
or $m_\nu/M_W$ where $\omega$ is the photon energy, $m_\nu$ is a neutrino
mass and $M_W$ is the $W$-boson mass.

Estimates\,\cite{bocc} have been made as to how much this  suppression in the
interaction would reduce the cross sections for neutrino photon scattering
but a precise calculation in the standard model seems not to have been carried
out.  The one--loop diagrams for neutrino photon scattering are shown in
Fig. 1.  In this paper, we perform the calculation in the zero
neutrino mass limit, assuming that the energy is much smaller than the
mass of the charged lepton.

\noindent{\large\sc 2.~Calculations}

A recent paper by Liu\,\cite{liu} considers non-standard interactions which
could give
non-zero neutrino -- two photon interactions at order $O(\alpha G_F)$ where
$\alpha$ is the fine structure constant and $G_F$ is the Fermi coupling.  We
consider only standard model interactions and calculate to order
$O(G_F^{\,2})$.

The calculation of the diagrams in Fig. 1 is simplified by the following
observations.
\begin{enumerate}
\item The matrix element for each diagram is of the form
$$
\bar{u}(p)(1+\gamma_5)\Gamma\,(1-\gamma_5)\,v(p')
$$
where $\bar{u}$ and $v$ are the spinors of the massless neutrinos and $\Gamma$
is
the appropriate combination of $\gamma$ matrices for the particular diagram.
This can be Fierz rearranged into
$$
{1\over 2} \;\bar{u}(p)\gamma^\lambda(1-\gamma_5) v(p')\,{\rm Tr}\,
\biggl(\Gamma\,\gamma_\lambda(1+\gamma_5)\biggr)
$$
and the trace can then be carried out using some
algebraic manipulation software.  (We used FORM \cite{form} and
SCHOONSCHIP \cite{schip}.)

\item A non-linear $R_\xi$ gauge condition can be chosen
such that the $\gamma W\phi$ coupling is zero\,\cite{gauge},
where $\phi$ is the Goldstone boson part of the $W$.  This gauge condition also
modifies the $\gamma\gamma WW$ and $\gamma WW$ couplings but the rules are
especially simple in the t'Hooft-Feynman gauge $\xi=1$.  We need calculate only
the diagrams in Fig. 1 plus those with every $W$ replaced by $\phi$.  Further,
the two contact diagrams (Fig. 1(d)) are zero. This choice of gauge also makes
it possible to check gauge invariance of the photon couplings analytically
rather than relying on a numerical check. For the case of zero neutrino mass,
the $W$--exchange and $\phi$--exchange diagrams are separately gauge invariant.
\end{enumerate}

After performing the loop integrals associated with each of the diagrams in
Fig.\,1, the remaining parameter integrals were expanded in powers of the
square
of the center of mass energy $s$ divided by a polynomial in the charged lepton
mass squared, $m_{\ell}^2$, and $M_W^2$. All parameter integrals are easily
evaluated and the $W$--exchange amplitude $\bigl({\cal M}_W\bigr)_{\lambda\,
\lambda^\prime}$, where $\lambda$ and $\lambda^\prime$ are the photon
helicities, is
\begin{equation}
\bigl({\cal M}_W\bigr)_{\lambda\,\lambda^\prime}  = \; {\sin\theta\over
M_W^{\,4}}\biggl(1 + \frac{4}{3}\ln({M_W^{\,2}\over m_\ell^2})\biggr)
\biggl(s\,t(1 - \lambda\lambda^\prime) + \frac{s^2}{2}(1 -
\lambda)(1 + \lambda^\prime)\biggr)\;. \label{gamgam}
\end{equation}
Here, $\theta$ is the scattering angle in the center of mass and only the
photon
helicities need be specified since the neutrino helicities are fixed. Notice
that only $\bigl({\cal M}_W\bigr)_{+\,-}$ and $\bigl({\cal M}_W\bigr)_{-\,+}$
are non--vanishing. The $W$--exchange amplitudes contain no inverse powers of
$m_{\ell}^2$ and are therefore of order $G_F^{\,2}$, as mentioned in the
Introduction. Explicit calculation shows that the $\phi$--exchange amplitudes
share this property. Since the $\phi$--lepton couplings introduce a factor
$m_{\ell}^2/M_W^{\,2}$, these amplitudes can be neglected to leading order.

The cross section for $\gamma\gamma\rightarrow  \nu\bar{\nu}$ is given by
\begin{equation}
\frac{d\sigma}{dz} = \frac{\sin^4\theta_W}{2^{13}\pi^5}
\;\frac{(M_W^{\,2}G_F)^4}{\omega^2}\;\sum\,|\bigl({\cal M}_W\bigr)_{\lambda\,
\lambda^\prime}|^2\;,\label{dsig}
\end{equation}
where $z$ is the cosine of the scattering angle, $\theta_W$ is the weak
mixing angle and $\omega$ is the center of mass energy of the photon.
The sum of the squared matrix elements is given by
\begin{equation}
\sum\,|\bigl({\cal M}_W\bigr)_{\lambda\,\lambda^\prime}|^2 =
\frac{2^4}{M_W^{\,8}}\,\biggl(1 + \frac{4}{3}\ln({M_W^{\,2}\over m_\ell^2})
\biggr)^2\biggl(-t^3\,(s + t) -t\,(s + t)^3\biggr)\;, \label{msq}
\end{equation}
with $s=4\omega^2\,$ and $t=-2\omega^2(1-z)$.  $m_\ell$ is the mass of
the charged lepton ($e,\mu,$ or $\tau$).
The presence of the logarithm in the matrix element makes the cross section
dependent on the flavor of neutrino.

The amplitudes for $\gamma\nu\rightarrow \gamma \nu$
and $\gamma\bar{\nu} \rightarrow  \gamma\bar{\nu}$ can be obtained from
eq.\,(\ref{gamgam}) with the appropriate interchange of $s,t,$ and $u$. For the
channel $\gamma\nu\rightarrow \gamma \nu$, the resulting amplitude
$\bigl(\tilde{\cal M}_W\bigr)_{\lambda\,\lambda^\prime}$ is
\begin{equation}
\bigl(\tilde{\cal M}_W\bigr)_{\lambda\,\lambda^\prime}  = \;
-2i{\cos(\theta/2)\over M_W^{\,4}}
\biggl(1 + \frac{4}{3}\ln({M_W^{\,2}\over m_\ell^2})\biggr)
\biggl(s^2(1 + \lambda\lambda^\prime) + \frac{s\,t}{2}(1 +
\lambda)(1 + \lambda^\prime)\biggr)\;, \label{gamnu}
\end{equation}
with $\theta$ denoting the scattering angle in the $\gamma\nu$ center of mass.
In this case, $\bigl(\tilde{\cal M}_W\bigr)_{++}$ and
$\bigl(\tilde{\cal M}_W\bigr)_{--}$ are non--vanishing. The sum of the squared
matrix elements is given by eq.\,(\ref{msq}) with $s$ and $t$ interchanged and
an overall minus sign. Because interaction is parity violating, the scattered
photons can acquire circular polarization. From eq.\,(\ref{gamnu}), the
polarization ${\cal P}(\theta)$ is
\begin{equation}
{\cal P}(\theta) = \,{|\bigl(\tilde{\cal M}_W\bigr)_{++}|^2 -
                      |\bigl(\tilde{\cal M}_W\bigr)_{--}|^2\over
                      |\bigl(\tilde{\cal M}_W\bigr)_{++}|^2 +
                      |\bigl(\tilde{\cal M}_W\bigr)_{--}|^2}\;\;
                 = \,{\cos^4(\theta/2) - 1\over
                       \cos^4(\theta/2) + 1}\;.
\end{equation}
Since there is no scattering in the Born approximation, polarization is of
order unity. Furthermore, the absence of any photon helicity flip amplitude
means that the scattering cannot produce linearly polarized photons.
${\cal P}(\theta)$ is plotted in Fig. 2.

\noindent{\large \sc 3.~Discussion}

The total cross sections for $\gamma\gamma\rightarrow\nu\bar{\nu}$ are obtained
from eq.\,(\ref{dsig}) using eq.\,(\ref{gamgam}), and those for $\gamma\nu$
elastic scattering are obtained using eq.(\ref{gamnu}) together with the
appropriate spin average. For electron neutrinos, the results are
\begin{eqnarray*}
\sigma_{\gamma\gamma\rightarrow  \nu\bar{\nu}} &=& 7.38 \times 10^{-50}\left(
\frac{\omega}{m_p}\right)^6\;\;{\rm cm}^2  \\
\sigma_{\gamma \nu\rightarrow \gamma \nu} &=& 1.11\times10^{-48} \left(
\frac{\omega}{m_p}\right)^6\;\;{\rm cm}^2 \;,
\end{eqnarray*}
where $m_p$ is the proton mass.  In each case $\sigma_{\gamma\bar{\nu}
\rightarrow \gamma\bar{\nu}}$ equals $\sigma_{\gamma \nu\rightarrow \gamma \nu}
$. The cross sections for all lepton species in units of $(\omega/m_p)^6$
cm$^2$
are summarized in Table 1. For comparison
\begin{table}[h]
\begin{center}
\caption{$\gamma\nu$ cross sections in units of $(\omega/m_p)^6\,$cm$^2$.}
\vspace{3pt}
\begin{tabular}{|c|c|c|}\hline
\mystr & $\sigma_{\gamma\gamma\rightarrow\nu\bar\nu}$ &
$\sigma_{\gamma\nu\rightarrow\gamma\nu}$\\
\hline
e\mystr       & 7.38$\times 10^{-50}$& 1.11$\times 10^{-48}$ \\
$\mu$\mystr   & 2.38$\times 10^{-50}$& 3.56$\times 10^{-49}$ \\
$\tau$\mystr  & 8.47$\times 10^{-51}$& 1.27$\times 10^{-49}$ \\ \hline
\end{tabular}
\end{center}
\end{table}
the cross section for
$\nu_e \nu_\mu\rightarrow  \nu_e\nu_\mu$ is $7\times 10^{-39}\;(\omega/m_p)^2$
cm$^2$.

The above expressions are valid only for energies $\omega\ll m_\ell$ as well
as $\omega\ll M_W$.  These limits cover most astrophysical applications and
mean that the standard model cross sections are negligibly small. Since
non--standard model amplitudes can be of order $\alpha G_F\,$\cite{liu}, it is
possible that the cross sections are larger in supersymmetric (SUSY)
models. The existence of a characteristic signature such
as circular polarization which could be attributed to the interaction of light
with the SUSY candidates for dark matter is worth further investigation.

\noindent{\large\sc Acknowledgements}

Palash Pal made two great suggestions.  First he reminded us of the non-linear
gauge conditions and second he pointed out that we could check our matrix
element in the forward direction with Ref.\,\cite{pall}.  We are very grateful
for his help. One of us (W.W.R.) wishes to thank G. Kane and B. Kayser for
helpful conversations. This research was supported in part by the National
Science Foundation under Grant 90-06117 and by the U.S. Department of Energy
under Contract No. DE-FG02-85ER40200.

\newpage


\pagebreak

\noindent
\centerline{\sc Figures}
\bigskip
\begin{figure}[h]
\hspace{1.5in}
\epsfysize=2.5in \epsffile{fig1.eps}
\caption{Diagrams for the process $\gamma\gamma\rightarrow  \nu_e\bar{\nu}_e$.
For each of (a), (b), (c) there are  also have the diagrams with the photons
interchanged.}
\end{figure}
\vspace{.2truein}
\begin{figure}[h]
\hspace{1.5in}
\epsfysize=2.5in \epsffile{fig2.eps}
\caption{Photon polarization for the process $\gamma\nu\rightarrow\gamma\nu$.}
\end{figure}

\end{document}